# Pinched Hysteresis Loops is the Fingerprint of Memristive Devices

Hyongsuk Kim, Maheshwar Pd. Sah, and Shyam Prasad Adhikari

*Abstract:* **This short note clarifies that the "pinched hysteresis loop" fingerprint of a memristor, or a memristive device, must hold for all amplitudes, for all frequencies, and for all initial conditions, of any periodic testing waveform, such as sinusoidal or triangular signals, which assumes both positive and negative values over each period of the waveform. We proved that the systems presented in [1] are *not* memristive devices because their hysteresis loops are *not* pinched at the origin for all amplitudes, and for all initial conditions**.

## 1. Introduction

A recent article [1] had suggested erroneously, via examples, that there are hypothetical memristive devices which exhibit a pinched hysteresis loop in the v-i plane which were not described by Eq. (1) defining a memristive device in [2]. The purpose of this note is to point out that Mouttet [1] had missed a crucial point in Chua's tutorial [3] which implies that the "pinched hysteresis loop fingerprint" corresponding to a sinusoidal input signal $i(t) = A \sin \omega t$ must be "pinched" at the origin, for *any* amplitude $A$, and for *any* frequency $\omega$, as well as for *any* initial condition $x(0)$ of the state variables at t = 0. Indeed, we quote the sentence above Eq. (26) of [3] as follow:

> "The loci (Lissajous) in the v-i plane of any *passive memristor* with positive memristance
> $$R(q) = \frac{d\hat{\varphi}(q)}{dq} > 0 \qquad (26)$$
> and driven by a sinusoidal current source $i(t) = A \sin \omega t$ is always a *pinched hysteresis loop, .....*"

The adjective "always" in the above sentence implies that the hysteresis loop must pass through the point v = 0 and i = 0 for any possible amplitude $A$, any initial state $x(0)$, and any possible frequency $\omega$, of the sinusoidal input signal $A \sin \omega t$.

We also note that the term "pinched" is defined in the Figure Caption (b) of Fig. 1 (page 767) of [3] which we reproduce as follow:

> (b) Pinched hysteresis loop: double-valued Lissajous figure of $(v(t), i(t))$ for all times *t*, except when it passes through the origin, where the loop is pinched.

In the following section, we will use the same examples presented in [1] to demonstrate that the "pinched hysteresis loop" presented in Mouttet's paper is a contrived "degenerate" case, in the sense that there exist other v-i hysteresis loops which are "not" pinched at the origin, when the amplitude A is different from "one", or when the initial condition x(0) is different from "zero", or when the testing signal is not a pure sinusoidal signal.

## 2. Examples Showing Hysteresis Loop is Not Pinched at the Origin When A ≠ 1

The $(v(t), i(t))$ loci (Lissajous figure) of a memristor, or memristive device, must pass through the *origin* for any magnitude A of the sinusoidal input. Fig. 1 (b) shows an example of 2 hysteresis loops which are *not* pinched at the origin; namely when A=1.5 (green hysteresis loop) and A=2(red hysteresis loop) respectively at x(0)= -A, . In Fig. 1(b), we apply the inputs signal $u(t) = A \sin \omega t$ to the system described by

$$y = u + \left(1 - (x)^2\right)x$$
$$\frac{dx}{dt} = \omega u \qquad (1)$$

which is the same system as Example 3 in [1].

_____________________________
H. Kim, M. Sah and S. P. Adhikari
Division of Electronics Engineering,
Chonbuk National University, Korea

In contrast, when the same input signal $u(t) = A\sin\omega t$ is applied to a memritive device described by[1]

$$M(t) = R_{ON}(x(t) + R_{OFF}(1 - x(t)))$$
$$\frac{dx(t)}{dt} = K\,i(t) \quad (2)$$

with K=10$^4$, $R_{ON}$ =100 Ohm, and $R_{OFF}$ = 16 KOhm, the device exhibits a hysteresis loops which is pinched at the origin, regardless of the amplitude A as depicted in Fig. 1 (a) with x(0)=0.

**3. Examples Showing Hysteresis Loop is Not Pinched at the Origin for Other Periodic Signals**

The pinched hysteresis loop property of a memristive device must also hold for *any* non-sinusoidal periodic waveforms $u(t)$ which assume both positive and negative values. The input u(t) signal in Fig. 2(a) is an example of a non-sinusoidal periodic waveform. This waveform is defined by, $u = A\sin\omega t + B\cos 2\omega t$ where A=0.8, B=0.8 and ω=3. Observe that the *v-i* loci of the memristor defined by (2) exhibits a pinched hysteresis loop at the origin. Fig. 2 (c) shows the *v-i* loci when the same input signal $u = A\sin\omega t + B\cos 2\omega t$ with $\frac{dx(0)}{dt} = -A\omega,\ x(0) = -\frac{B}{4}$ is applied to the system described by

$$y = u + x\frac{du}{dt}$$
$$\frac{d^2x}{dt^2} = \omega^2 u \quad (3)$$

which is the same system as Example 2 of [1]. Observe that the *v-i* loci shown in Fig. 2(c) exhibits 2 points where $y \neq 0$ on the vertical axis (u=0). This is not possible for a *passive* memristor. Moreover since the *v-i* loci wanders into the second quadrant, and the fourth quadrant, this system is *not* passive, and can never be used as a non-volatile memory.

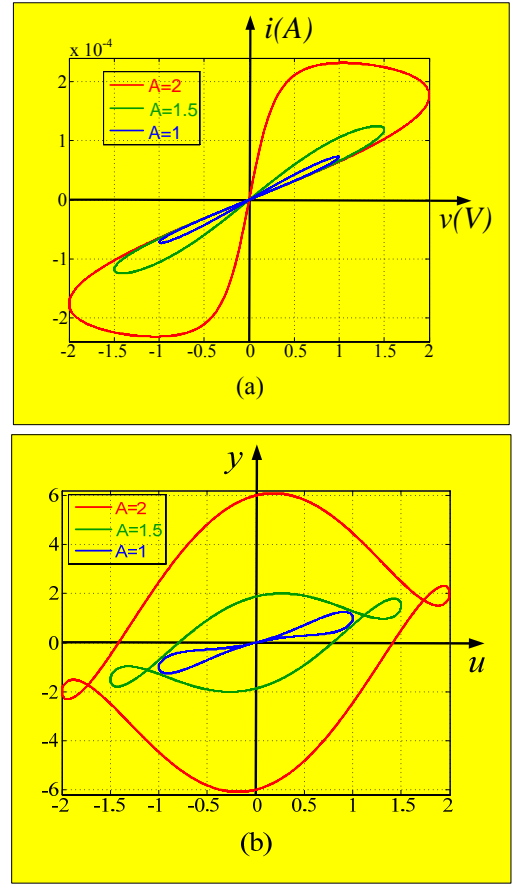

Fig. 1. A comparison between a memristive system and a non-memristive system when the amplitude A of the input signal $u(t) = A\sin\omega t$ (ω=5) is not equal to A=1. (a) All *v-i* loci are hysteresis loops pinched at the origin. (b) Hysteresis loop at A=1.5(green) and A=2(red) are not pinched at the origin.

The same simulation has been conducted with another *non-sinusoidal* period input signal $u = A\sin(\omega t) + B\cos(6\omega t)$ (with $\frac{dx(0)}{dt} = -A\omega$, $x(0) = -\frac{B}{36}$ and A=0.8, B=0.8, and w=3), as shown in Fig. 3(a). Observe that the hysteresis loop in Fig. 3(b) corresponding to the memristor described by Eq. (2) is pinched at the origin.

---

[1] Eq. (2) is a simplified version of the hp memristor equation [4]

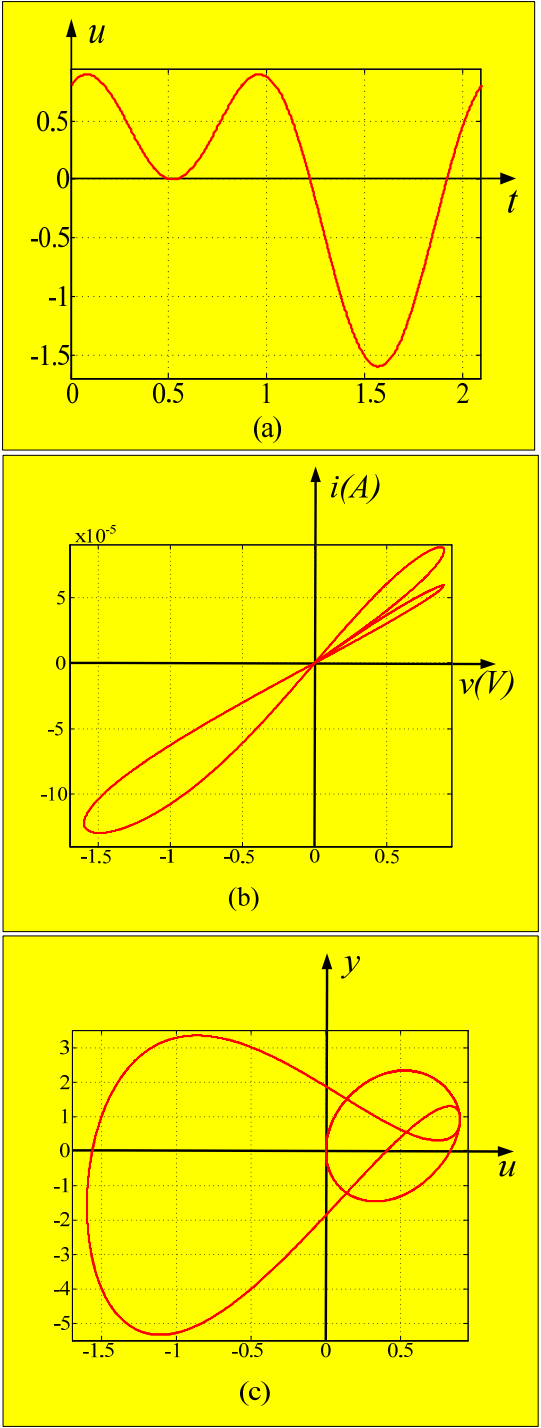
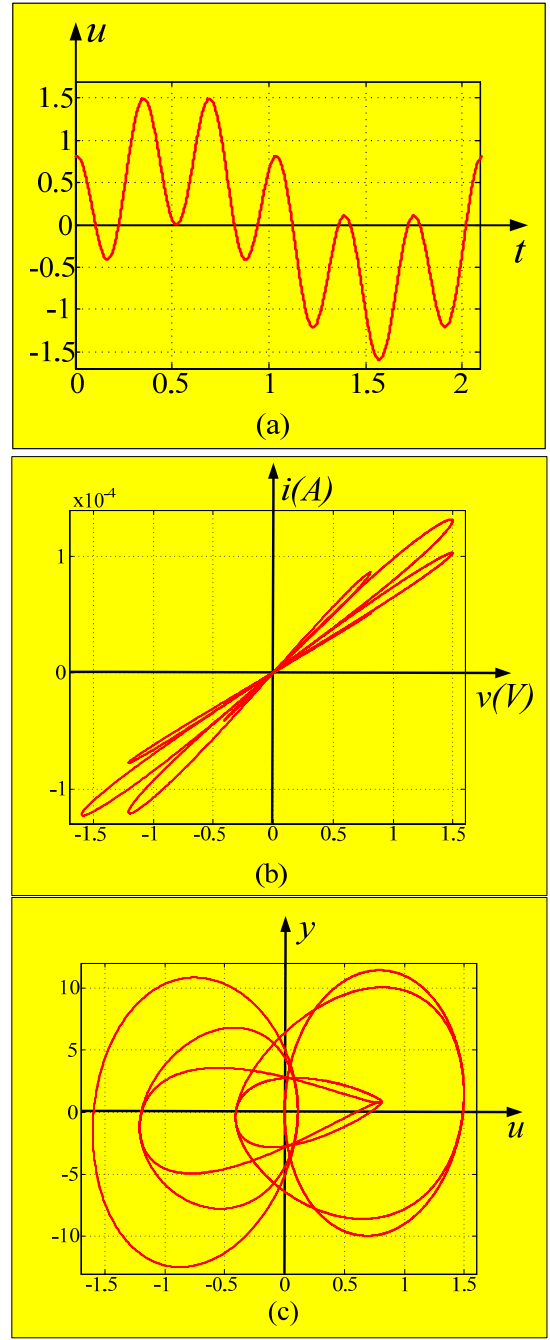

Fig. 2. Comparison of v-i loci when a *non-sinusoidal* input signal $u = A\sin\omega t + B\cos 2\omega t$, with A=0.8, B=0.8 and ω=3, is applied to a memristive system and a non-memristive system. (a) Input signal (single period). (b) Pinched hysteresis loop obtained from the memristive system described by Eq. (2). (c) The hysteresis loop obtained from the non-memristive system described by Eq. (3) contains 2 points $(0, y_1)$ and $(0, y_2)$ where $y_1 \neq 0$ and $y_2 \neq 0$.

Fig. 3. Comparison of v-i loci when *a non-sinusoidal* input signal $u = A\sin(\omega t) + B\cos(6\omega t)$, where A=0.8, B=0.8 and ω=3, is applied to both a memristive system and a non-memristive system. (a) Input signal (single period). (b) Pinched hysteresis loop obtained from the memristive system described by Eq. (2). (c) The hysteresis loop obtained from the non-memristive system of Eq. (3) contains many points with $y \neq 0$ on the u=0 axis. Observe that $u(t) \times y(t) < 0$, whenever the loci wanders into the 2$^{nd}$ and the 4$^{th}$ quadrants. This means that the device described by Eq. (3) is delivering power to the external circuit, and is therefore an active device

requiring an internal power supply.

Another simulation of system (3) with a *non-sinusoidal* period input signal $u = A\sin(\omega t) + B\sin(2\omega t) + C\cos(\omega t) + D\cos(2\omega t)$ (with $\frac{dx(0)}{dt} = -\omega\left(A + \frac{B}{2}\right)$, $x(0) = -\left(C + \frac{D}{4}\right)$, and A=0.8, B=0.8, C=0.8, D=0.8 and ω=3) is shown in Fig. 4. Observe that the *v-i* loci of the memristor described by Eq. (2) is shown in Fig. 4(b). Note that even though the hysteresis loop in this case has multiple lobes, it is pinched at the origin, and its excursion is restricted within the first and third quadrants only, a necessary condition for passivity.

In contrast, the *v-i* loci in Fig. 4(c) corresponding to the system (3) is not pinched at the origin. Moreover, the loci wanders into the second and fourth quadrants where y(t)×u(t)<0, implying that to build a device described by (3), a power supply must be imbedded within the device, namely, the device is not passive.

## 4. Examples Showing Hysteresis Loop is Not Pinched at the Origin When Initial States x(0) is Non-Zero

A memristor or a memrisitve device driven by periodic input which assumes both positive and negative values, must always exhibit a hysteresis loop in steady state (i.e. after the transients have decayed to zero) which is *pinched* at the origin for *any initial* condition x(0). Fig. 5 (a), (b), and (c) shows the *v-i* loci when u=sin ωt is applied to the system described by Eq. (1), Eq. (3), and Eq. (4) (below), obtained from Example 1 of [1].

$$y = u + 2x$$
$$\frac{dx}{dt} = u^2 - \frac{1}{2} \qquad (4)$$

Observe the loci depicted in Fig. 5(a), (b) and (c) are not pinched at the origin. In contrast, when the same input signal u= sin ωt , ω=10 is applied to the memristive system in Eq (2), a pinched hysteresis loop obtained at all initial states, as shown in Fig. 6.

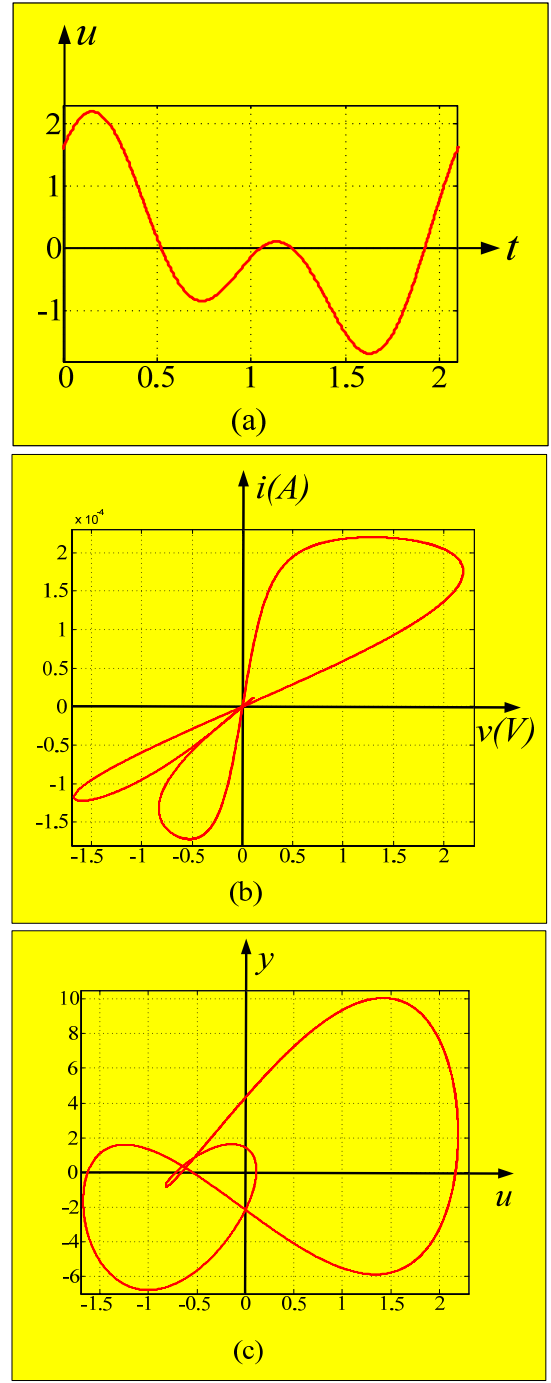

Fig. 4. Comparison of v-i loci when a *non-sinusoidal* periodic input signal $u = A\sin(\omega t) + B\sin(2\omega t) + C\cos(\omega t) + D\cos(2\omega t)$ where A=0.8, B=0.8, C=0.8, D=0.8, and ω=3, is applied to a memristive system and a non-memristive system. (a) Input signal (single period). (b) Pinched hysteresis loop obtained from the memristive system described by Eq. (2) (c) Non-pinched hysteresis loop obtained from the non-memristive system described by Eq. (3).

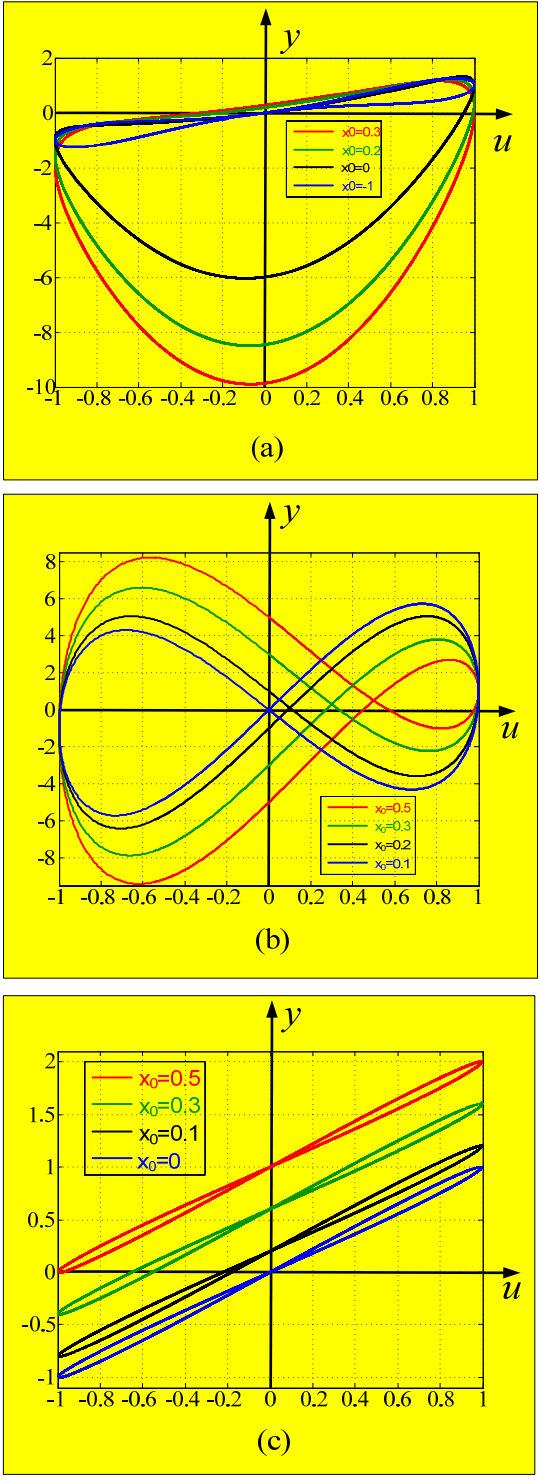

Fig. 5. Hysteresis loops of non-memristive systems with various *non-zero* initial states. (a) System described by Eq. (1). (b) System described by Eq. (3), and (c) System described by Eq. (4). In this case, all hysteresis loops with x(0)≠0 can be easily proved to be a vertical translation of the blue hysteresis loop corresponding to x(0)=0, and hence are not pinched at the origin (u,y)=(0,0).

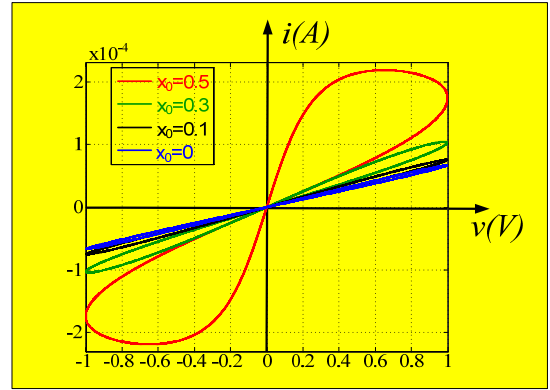

Fig. 6. Pinched hysteresis loops of the memristive systems described by Eq. (2), with various non-zero initial states

## 5. Concluding remarks

The pinched hysteresis loop is indeed the fingerprint of memristor and holds true for any amplitude and frequency of the input sinusoidal, as well as for any valid initial condition of the memristor's state variable. The v-i loop for memristor is also pinched at the origin for any non-sinusiodal periodic waveform. It follows from the above examples that the various assertions in [1] are incorrect and are at best misleading. We also take this opportunity to point out that a similar-sounding 3-terminal device called the memistor, which was asserted to be a memristor [5], is in fact different from a memristor [6],[7].